\documentclass[twocolumn]{svjour3}

\usepackage{graphicx} 
\usepackage{amsmath,amssymb}
\usepackage{footnote}
\usepackage{hyperref}
\usepackage{siunitx}
\usepackage{xspace}
\def\MADMAX{\mbox{{\sc Madmax }}}
\def\ADMX{\mbox{{\sc Admx }}}
\def\ORGAN{\mbox{{\sc Organ }}}
\def\HAYSTAC{\mbox{{\sc Haystac }}}
\def\CULTASK{\mbox{{\sc Cultask }}}

\newcommand{\FrequencyRange}{\SIrange{10}{100}{\giga\hertz}\xspace}

\newcommand{\orcid}[1]{\href{https://orcid.org/#1}{\includegraphics[width=10pt]{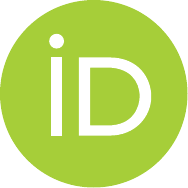}}}
\usepackage{hyperref}

\usepackage{etoolbox}
\apptocmd{\sloppy}{\hbadness 1000\relax}{}{}

\journalname{Computing and Software for Big Science}
\begin{document}
\sloppy

\title{Simulation of dielectric axion haloscopes with deep neural networks: a proof-of-principle}

\author{Philipp~Alexander~Jung\textsuperscript{1}~\orcid{0000-0002-2511-1490} \and     
        Bernardo~Ary~dos~Santos\textsuperscript{1}~\orcid{0000-0003-0796-7524} \and 
        Dominik~Bergermann\textsuperscript{1}~\orcid{0000-0002-8738-6289}  \and 
        Tim~Graulich\textsuperscript{1}~\orcid{0000-0002-7442-723X} \and 
        Maximilian~Lohmann\textsuperscript{1}\orcid{0000-0003-3476-5856}\and 
        Andrzej~Nov\'{a}k\textsuperscript{1}~\orcid{0000-0002-0389-5896} \and 
        Erdem~\"Oz\textsuperscript{1}\orcid{0000-0003-2621-4990}\and 
        Ali~Riahinia\textsuperscript{1}~\orcid{0000-0003-2736-1861} \and 
        Alexander~Schmidt\textsuperscript{1}~\orcid{0000-0003-2711-8984}
}
\authorrunning{Philipp~Alexander~Jung et. al.}

\institute{Correspondence to: novak@physik.rwth-aachen.de\at \\   
\textsuperscript{1} RWTH Aachen University, Aachen, Germany\\
}
\date{\ }

\maketitle
\begin{abstract}
Dielectric axion haloscopes, such as the \MADMAX
experiment, are promising concepts for the direct search for dark matter axions. A reliable simulation is a fundamental requirement for the successful realisation of the experiments. Due to the complexity of the simulations, the demands on computing resources can quickly become prohibitive. In this paper, we show for the first time that modern deep learning techniques can be applied to aid the simulation and optimisation of dielectric haloscopes. 
 
\end{abstract}

\section{Introduction}
Axions are hypothetical bosons predicted by the Peccei--Quinn (PQ) mechanism. The PQ mechanism explains the absence of CP-violation in quantum chromodynamics (QCD)~\cite{cite.physrevlett.38.1440,cite.physrevd.16.1791,cite.physrevlett.40.223,cite.physrevlett.40.279}. Axions could also be responsible for the cold dark matter (DM) of the universe~\cite{cite.preskill1983127,cite.abbott1983133,cite.dine1983137}. Therefore, the existence of axions would resolve two major problems in physics. Due to astrophysical constraints \cite{cite.raffelt:2006cw} the mass of the axion would have to be very small $m_a < \SI{20}{\milli\electronvolt}$ and various cosmological models constrain the mass range even further \cite{cite.kawasaki:2014sqa,cite.physrevd.85.105020,cite.physrevd.49.5040,cite.zurek:2006sy,cite.ballesteros:2016xej}. 

Microwave cavity haloscopes \cite{cite.physrevlett.51.1415,cite.physrevlett.52.695.2} such as \ADMX \cite{cite.physrevd.69.011101,cite.physrevlett.120.151301}, \ORGAN \cite{cite.mcallister:2017lkb}, \HAYSTAC~\cite{cite.haystac} and \CULTASK~\cite{cite.cultask} are one class of experiments searching for low-mass DM axions. The principle of these experiments is based on axion--photon conversion in the presence of a strong magnetic field. A new experimental design is a dielectric haloscope, which will be able to probe higher mass ranges and higher frequencies than the experiments mentioned above. The MAgnetized Disk and Mirror Axion eXperiment (\MADMAX) \cite{cite.brun:2019lyf} will be the first realisation of a dielectric haloscope with the goal of reaching the sensitivity for the QCD Axion.

\begin{figure}
    \centering
    \includegraphics[width=0.47\textwidth]{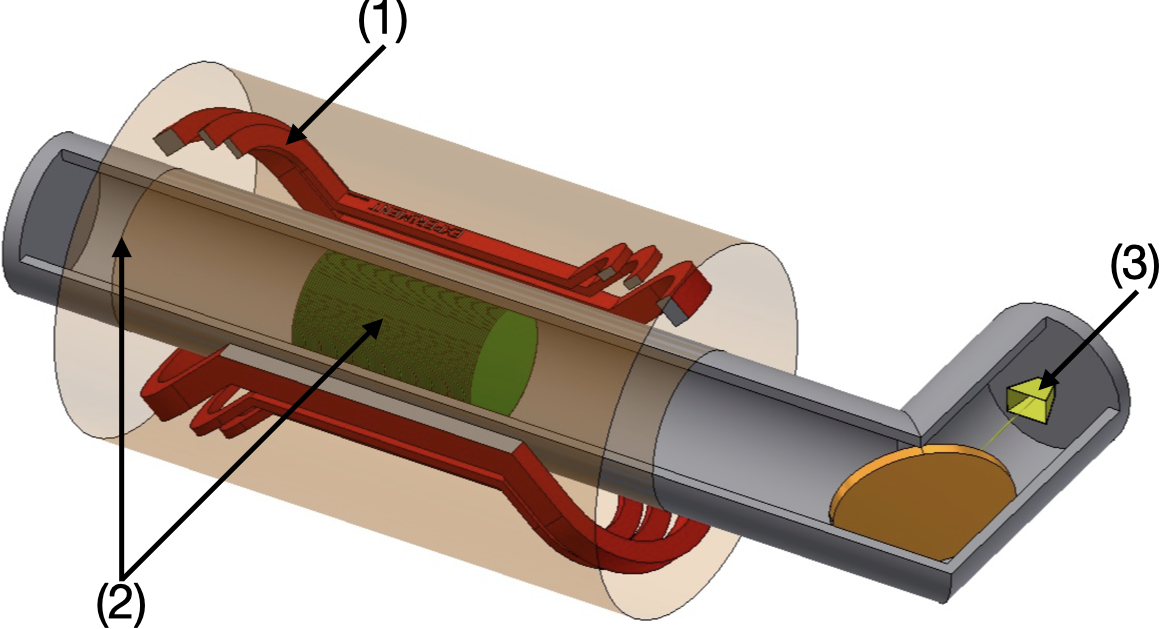}
    \vspace{20pt}
    \caption{Preliminary baseline design of the MADMAX approach. The experiment can be divided into three parts: (1) magnet (red racetracks), (2) booster – consisting of the mirror (copper disk at the very left), the 80 dielectric disks (green) and the system to adjust disk spacing (not shown) – (3) the receiver – consisting of the horn antenna (yellow) and the cold preamplifier inside a separated cryostat. The focusing mirror is shown as an orange disk at the right. Not to scale. Figure reproduced from Ref.~\cite{cite.brun:2019lyf}.}
    \label{fig.madmax}
\end{figure}

In a dielectric haloscope, microwave radiation is emitted from the surfaces of dielectric disks as a result of axion--photon conversion at the interface between the disk and the vacuum \cite{cite.millar:2016cjp}. The signal from a single interface would be too weak to detect, therefore the signal is amplified (boosted) through resonance and constructive interference from multiple disks. The enhancement compared to a single metal surface is the boost factor $\beta$. The frequency and bandwidth at which $\beta$ is maximal can be adjusted through disk placement. Even for fixed disk thicknesses, the variation of distances between disks allows one to scan a large frequency range. MADMAX aims at probing the frequency range of \FrequencyRange. A prototype of the MADMAX experiment will consist of 20 disks, while the final experiment will likely use around 100 disks. Figure~\ref{fig.madmax} shows a sketch of the planned experiment.

Simulations of the experiment can be used for sensitivity estimates as well as to constrain experimental boundary conditions such as the required mechanical precision of the disk placement system. Various approaches can be used for the simulation of the experiment at different levels of detail. So-called analytical 1D methods \cite{cite.millar:2016cjp} are used in an idealised context, where the disk size is taken to be infinite and some design aspects are neglected. Actual 3D simulations \cite{cite.knirck:2019eug} are more realistic but suffer from high demands on computing resources and quickly become prohibitive, especially when optimising in a 3D simulation with a large number of disks and therefore a large number of degrees of freedom. An example, where alternative simulation approaches may be helpful is the optimisation of the disk positioning for a desired frequency and bandwidth of the boost factor. Optimisation needs to be done iteratively by simulating the electromagnetic response of the experiment for a fixed set of disks many times. Disks are readjusted in each iteration until the desired boost factor curve is reached. 

Applications of gradient descent to find an optimal disk configurations have already been explored~\cite{McDonald_2022}. In this work, however, we demonstrate a proof-of-principle that machine learning methods are capable of reproducing the actual electromagnetic response of the experiment. Specifically, we train a neural network which predicts a boost factor curve from a set of disk positions.

\section{Simulation and dataset}
To generate the training data, the analytical 1D simulation is used to calculate the boost factor curves corresponding to various configurations of the disk positions. The 1D simulation assumes a perfect idealised geometry with 20 plane disks of infinite diameter and of $\SI{1}{\milli\meter}$ thickness and a mirror on one side of the cavity. The total length of the apparatus is variable but on the order of a few metres. The calculations are based on the modified axion--Maxwell's equations from which the axion-induced emission of electromagnetic radiation from an interface between the dielectric and vacuum is derived. Each interface has an incoming and an outgoing wave that satisfy the overall boundary conditions of the system. To calculate the resulting signal that emerges from the system a transfer matrix formalism \cite{cite.millar:2016cjp} is applied. 

The reference boost factor curves and their corresponding disk configurations for frequencies covering a range from $\SI{19}{\giga\hertz}$ to $\SI{25}{\giga\hertz}$ with a bandwidth of $\SI{50}{\mega\hertz}$ are found through iterative optimisation of the disk placement using the Nelder-Mead algorithm \cite{cite.nelder-mead}. The bandwidth of the curve is considered during the optimisation procedure, in which the target of the objective function is defined by the lowest value of the boost factor in the desired frequency window.

Generating a truly random dataset by varying the disk spacings without any constraints would currently be prohibitive, as the fraction of the ``experimentally interesting''  boost factor curves would be low. Experimentally interesting curves are box-like curves of non-negligible bandwidth in the desired frequency range. Instead, the pre-optimised reference configurations are used as seeds. For each seed configuration, the disk positions are randomly varied with a uniform distribution within a range corresponding to 5\% of the disk thickness --- $\SI{0.05}{\milli\meter}$. These pseudorandom curves are then filtered to have bandwidths (FWHM) between $\SI{40}{\mega\hertz}$ to $\SI{200}{\mega\hertz}$ and maxima of the boost factor between $15\,000$ to $25\,000$. This requirement rejects about 90\% of the generated configurations.

\begin{figure}[!htbp]
    \centering
    \includegraphics[width=0.47\textwidth]{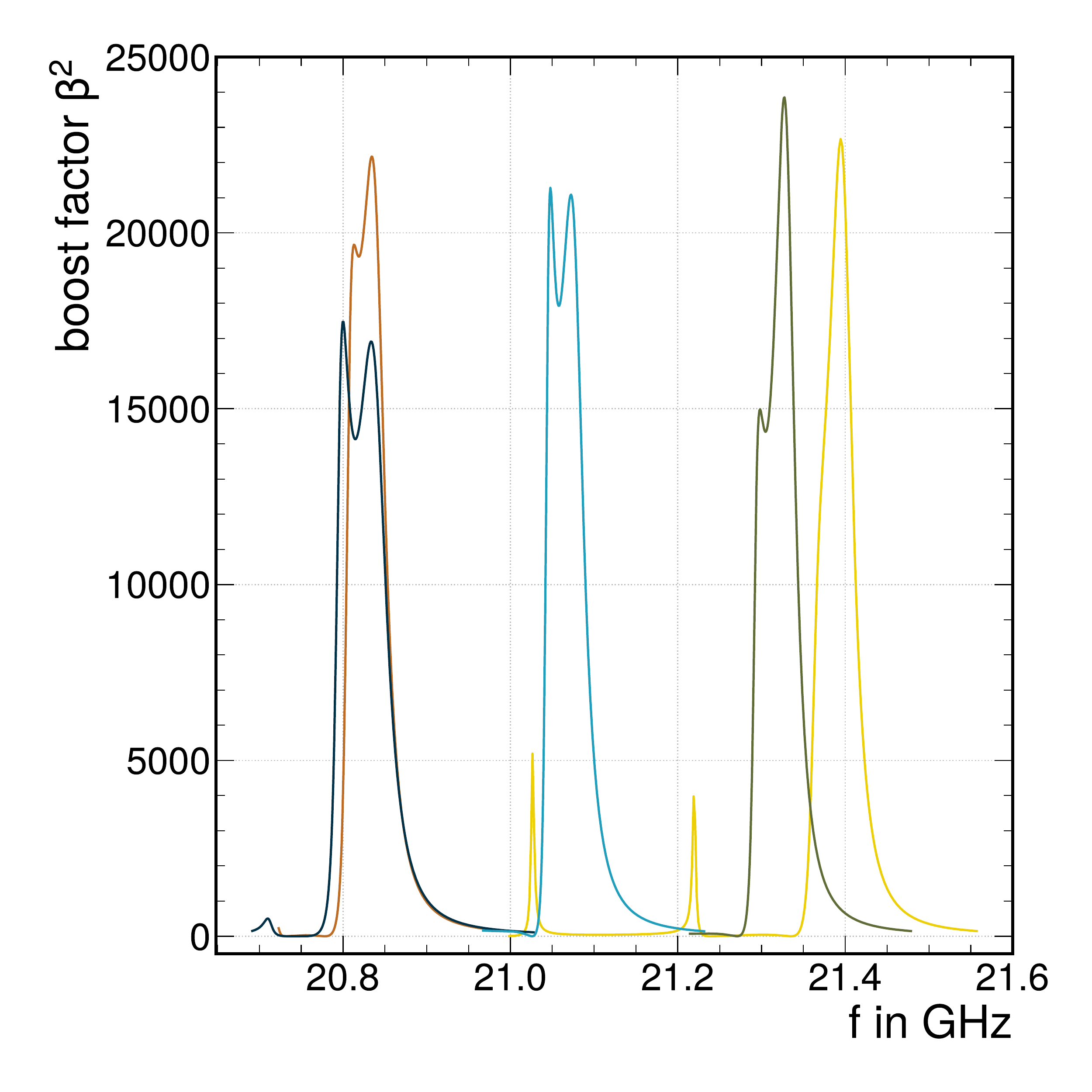}
    \caption{Nominal pre-optimised boost factor curve (light blue line) with a bandwidth of $\SI{50}{\mega\hertz}$. Also shown are several boost factor curves with randomly varied disk positions according to a uniform distribution with a maximum variation of 5\%.}
    \label{fig.boostvariation}
\end{figure}

In addition, the integral of the boost factor over the entire frequency space is constant by the area law, regardless of the disk positions \cite{cite.millar:2016cjp}. This property can be used to determine whether a configuration is sampled at the correct frequency space. Configurations with an integral of less than 95\% 
are discarded. 
Figure~\ref{fig.boostvariation} shows an optimised reference boost factor curve and several random curves generated according to the above prescription.
It is visible that even these small variations of the disk positions result in significant deformations of the boost factor curve.
\begin{figure*}[!htbp]
    \centering
    \includegraphics[width=1.0\textwidth]{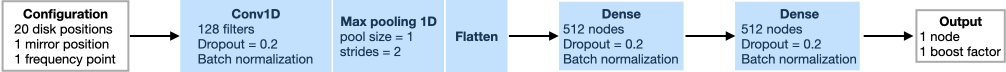}
    \caption{Model architecture. All layers except the last one use PReLU as the activation function with alpha initializer of 0.25.}
    \label{fig.network}
\end{figure*}
The final dataset contains $169\,260$ configurations and corresponding boost factor curves, sampled at 300 points each along the frequency space.  The dataset is further divided into training, validation and testing sets with $122\,290$, $21\,581$ and $25\,389$ configurations respectively.

\section{Network architecture and training}
\subsection{Inputs}
\label{sec.inputs}
In order to build a model capable of predictions at arbitrary frequency values the input data is reshaped such that each entry consists of an array of distances between the disks and a frequency value at which the boost factor is sampled. The corresponding truth target is then the boost factor value at that frequency. A single configuration consisting of 21 (20 disks, 1 mirror) positions with corresponding boost factor values at 300 frequency values becomes $22 \cdot 300 = 6600$ individual training data points.
To facilitate training, the input data and the truth target are transformed, respectively, into a uniform distribution using a Quantile Transformer \cite{cite.scikit}.

\subsection{Model and Training}
\label{sec.Model}
The model consists of a 1D convolutional layer~\cite{cite.cnn} with 128 filters with subsequent 1D max pooling, as well as flattening, followed by two dense layers with 512 nodes each and an output node. Each layer, except the last, uses batch normalisation \cite{cite.batchnorm}, dropout \cite{cite.dropout} at a rate of 0.2 and PReLU \cite{cite.prelu} as an activation function with an alpha initializer of 0.25, following the recommended initialisation of the weights using the He uniform initializer. The structure of the network is shown in Fig.~\ref{fig.network}.

The Adam optimiser \cite{cite.adam} is used for training in its default configuration with a learning rate of $10^{-5}$. Mean absolute error (MAE) serves as the loss function in addition to the L$^2$ norm with a weight of $10^{-4}$. Early stopping in combination with model checkpoints is used to prevent overtraining. The batch size is 1024.

The presented model is chosen based on a cursory search and evaluation of various architectures, including layer type, layer number, number of nodes, activation functions, cost function, weight initialisation, and preprocessing.

\subsubsection{Generalization across frequency phase space}
\label{sec.blinding}
In addition to the usual measures of controlling overtraining, such as the aforementioned regularisation and early stopping, it is of interest to see if the model can generalise reasonably well within the target phase space. The expectation is that the network should be able to predict the boost factor values at arbitrary frequency values within the target range, because the inputs are effectively smooth. To that end, a new training is conducted with a modified training set, where certain frequency ranges are blinded. Specifically, we remove frequency bands of $\SI{500}{\mega\hertz}$, around each half integer between $\SI{19}{\giga\hertz}$ and $\SI{25}{\giga\hertz}$, keeping about half of the original data for the training. The validation and test sets are kept the same as for the baseline training.
\begin{figure}[!htbp]
    \centering
    \includegraphics[width=0.47\textwidth]{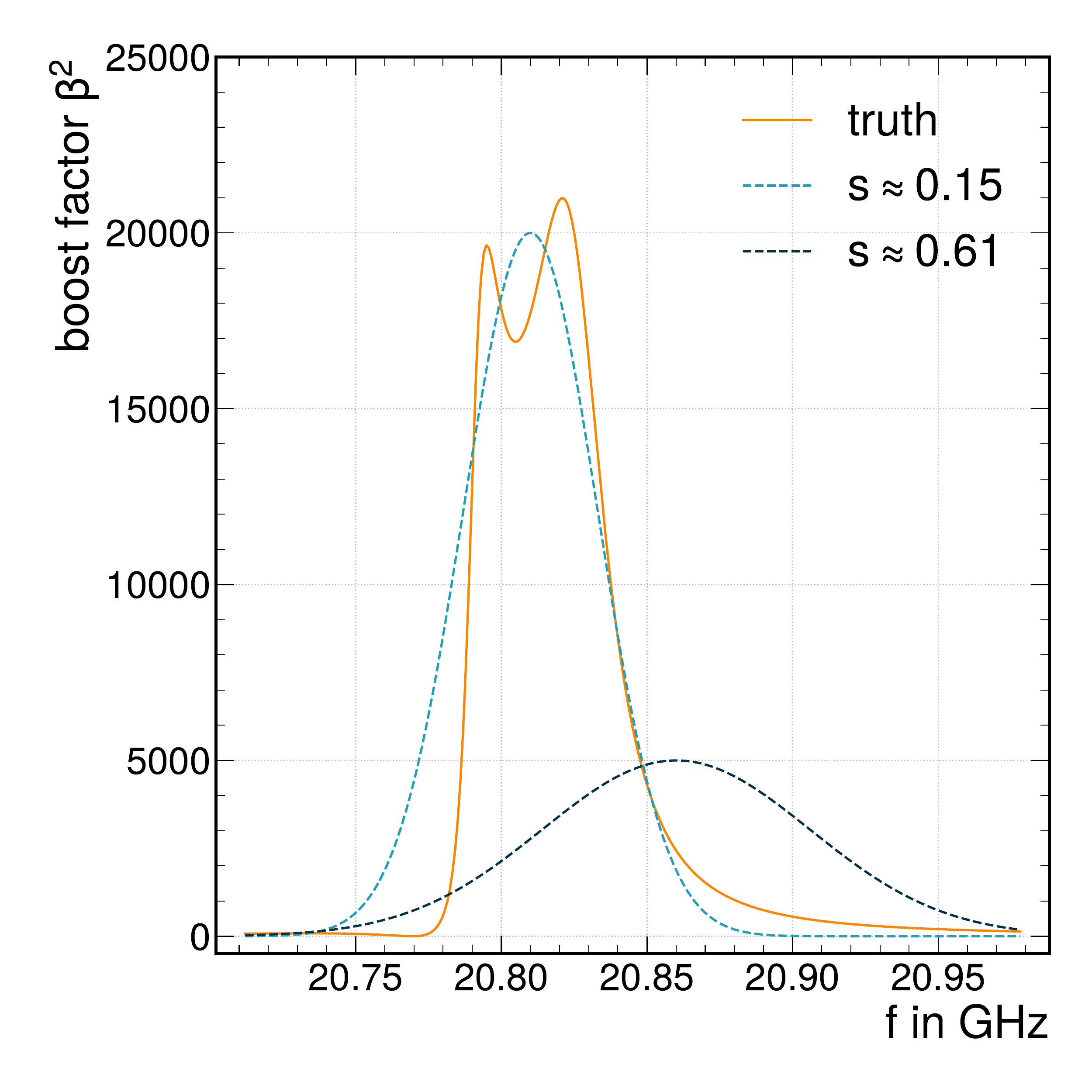}
    \caption{Example prediction with a ``good'' boost score of $s \approx 0.15$ (light blue) and a ``poor'' boost score of $s \approx 0.61.$ (dark blue) with the truth (orange).}
    \label{fig.metric}
\end{figure}
\section{Performance evaluation}
\subsection{Evaluating the prediction}
The prediction task is somewhat atypical, so common performance metrics such as mean absolute error (MAE)
\begin{align*}
    \text{MAE} = \text{mean}\left(\left| \sum_i y_{true}^i - y_{prediction}^i \right|\right)
\end{align*}
are not ideally suited to describing how well a predicted boost factor curve matches the truth. Important features of interest are peak positions and bandwidth. However, the peak position often cannot be unambiguously determined due to the typical double-peak structure, which can be seen in Fig.~\ref{fig.result_comparison}. Furthermore, we are interested in the overall shape rather than the accuracy at any specific point. We therefore define a new metric suitable for this task by comparing the overlapping areas of the curves. 

The metric is named boost score in the following and is defined as
\begin{align}
    s = 1 - \frac{2 \cdot A_o}{A_t + A_p},
    \label{eq.metric}
\end{align}
where the score $s$ is given by the area of the truth curve $A_t$, the area of the prediction curve $A_p$, and the overlapping area $A_o$. Thus, $s \in [0,1]$, where $0$ represents a perfect prediction and $1$ represents the largest possible deviation from the truth. 

Figure~\ref{fig.metric} illustrates the behaviour of the boost score metric and shows that the boost score responds both to the peak position and width intuitively, making it a suitable measure of performance for this problem. 

\section{Results}

To investigate the stability of the selected architecture, the trainings were repeated five times each. These 5 runs differ only in the random initialisation of the weights in the network. The constant behaviour of the runs indicates a good robustness of the chosen architecture. Small differences are to be expected due to the stochastic nature of the neural network.

\begin{figure}[!htbp]
    \centering
    \includegraphics[width=0.47\textwidth]{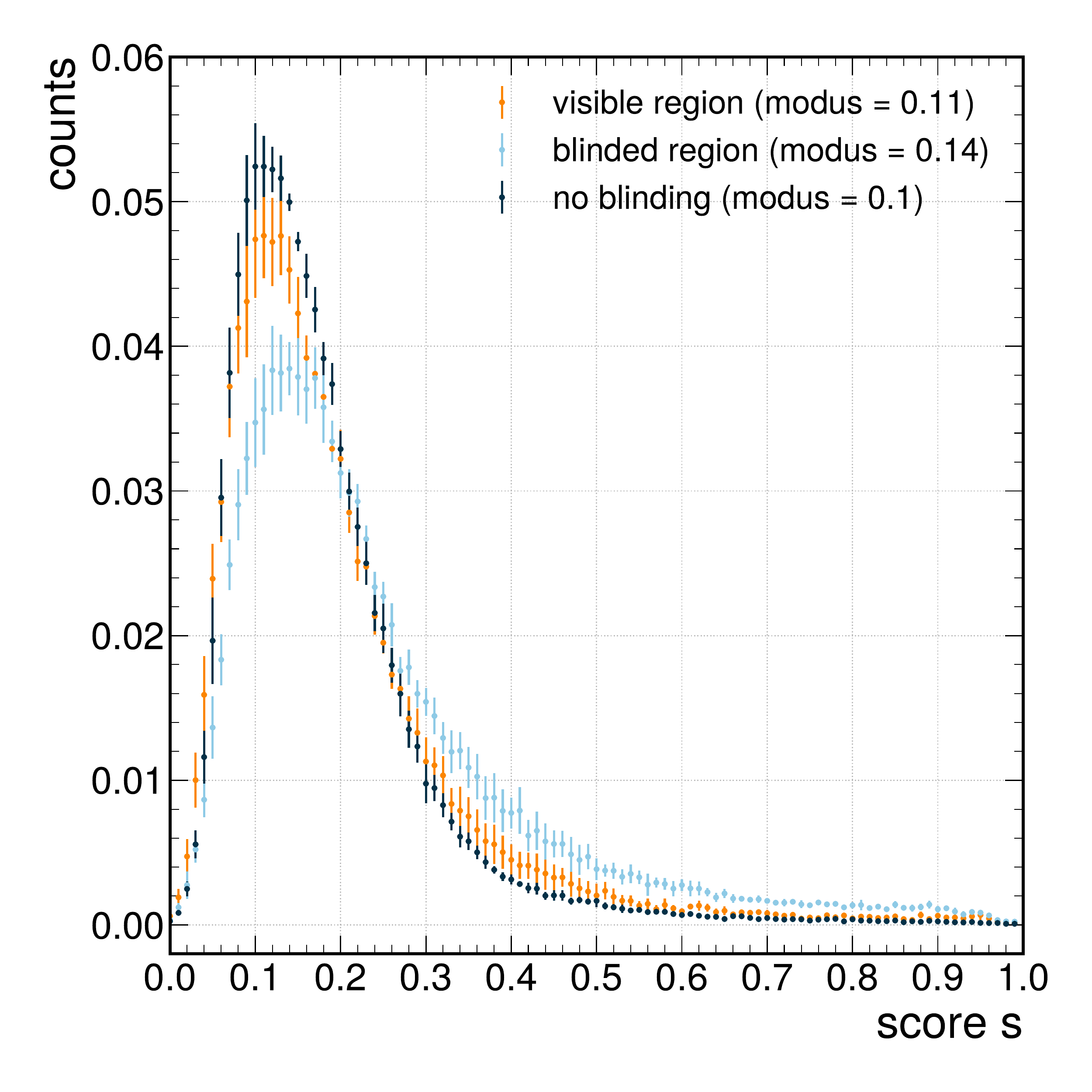}
    \caption{Boost score distribution averaged over 5 trainings of the baseline configuration (dark blue). As well as the distributions of the blinded training split between the blinded and visible frequency range.}
    \label{fig.result_comparison}
\end{figure}

The overall performance of the trained model is shown in Fig.~\ref{fig.result_comparison}, where the bin-wise mean score of the 5 runs with its respective standard error is shown. Empirically, a score below $\approx~0.3$, that is, covering $70\%$ of the expected boost factor area, is sufficiently accurate to be of experimental value for \MADMAX~boost factor predictions. For the baseline training, $89\%$ of the predictions reach this target. For the blinded training $74\%$ of the predictions in blinded frequency range and and $84\%$ of the predictions in the visible range meet this target.

\begin{figure*}
    \centering
    \includegraphics[width=1.0\textwidth]{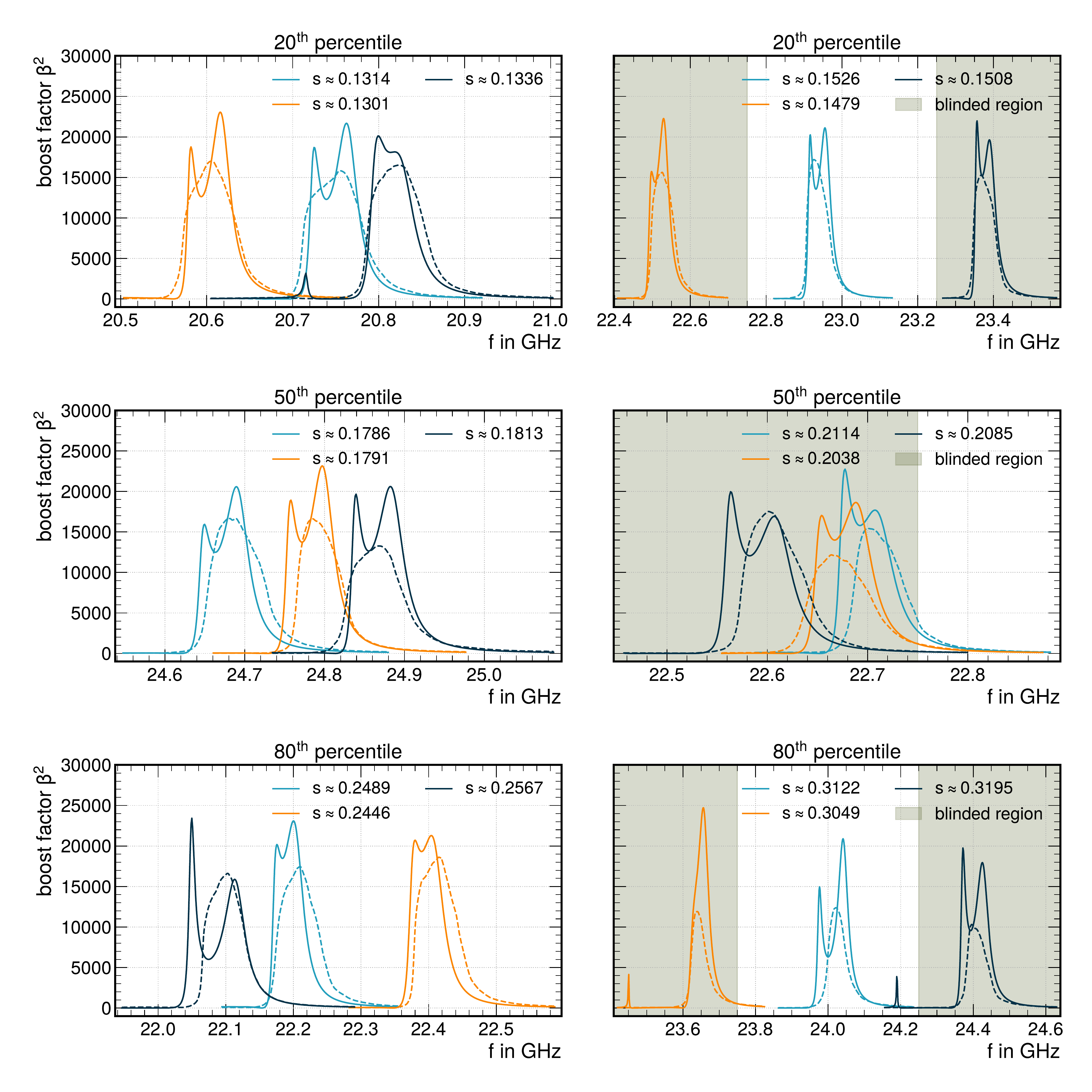}
    \caption{Several randomly selected predictions from the test set at the $20^\text{th}$, $50^\text{th}$ and $80^\text{th}$ percentiles are shown for the baseline (left) and blinded (right) trainings. The solid line represents the simulated ``truth'' boost factor curve and the dashed line the prediction of the network. The corresponding boost score is given. The blinded area is shown in green.}
    \label{fig.results}
\end{figure*}

In addition to provide a sense of the prediction quality, randomly sampled curves from the test set are shown in Fig.~\ref{fig.results} at several boost score percentiles for both the baseline (left) and blinded (right) trainings. \footnote{Percentiles are obtained form a sorted score distribution entries divided into 100 equal parts}.

In general the model can reproduce the peak position and the bandwidth sufficiently well. However, the actual peak structure is typically not captured. By comparing the baseline with the blinded training we show the model is able to interpolate across masked phase space, albeit at a certain loss to performance, indicating good generalization capabilities.

A major promise of using deep learning to estimate the boost factor is computational efficiency. The analytical 1D simulation takes about $\SI{80}{\second}$ to generate the curves from the test set, which translates to about $\approx320\,\text{curves/s}$. Corresponding values for a much more realistic three-dimensional simulation range from $\SI{1400}{\second}$ and $18\,\text{curves/s}$ to $\SI{80e6}{\second}$ and $0.0003\,\text{curves/s}$, depending on the desired accuracy. Compared to the simulation of the analytical 1D code, the network prediction is faster by about a factor of 5. Moreover so, when considering the more realistic 3D simulations, the speedup is on the order of millions to tenths of millions, depending on the selected precision. Meanwhile, the prediction speed of the network should be independent of the training data used.

\section{Conclusion}
In this paper we present a first application of deep learning methods to model the boost factor response of an axion haloscope experiment. Although the current implementation has some limitations imposed in terms of the phase space covered, it is capable of predictions accurate enough to be of experimental value, while also being very robust. Furthermore, we show that the model is able to predict into fully blinded frequency regions, demonstrating the ability to generalize, although the quality of the prediction suffers a bit.

\begin{acknowledgements}
Simulations were performed with computing resources granted by RWTH Aachen University under project rwth0583. We thank the Bundesministerium für Bildung und Forschung (BMBF) for the support under project numbers 05H20PARDA and 05H21PARD1.
\end{acknowledgements}

\section*{Ethics Declaration}
The authors declare that they have no conflict of interest.

\section*{Data Availability Statement}
The datasets generated and analysed during the current study are available from the corresponding author on reasonable request.

\bibliographystyle{lucas_unsrt}
\bibliography{bibliography}
\end{document}